\documentclass{appolb}

\usepackage{graphicx}
\usepackage{amssymb}
\usepackage{hyperref}
\usepackage[utf8]{inputenc}

\begin{document}

\title{Non-exponential decay in classical stochastic processes with memory%
\thanks{e-mails: Maciej.Rybczynski@ujk.edu.pl, Zbigniew.Wlodarczyk@ujk.edu.pl}}

\author{Maciej Rybczy\'nski and Zbigniew W\l odarczyk 
\address{Institute of Physics, Jan Kochanowski University, PL-25406~Kielce, Poland}
}

\maketitle

\begin{abstract}
The initial time-dependence of a state in circumstances where it makes transitions to, or decay to, a second state has been investigated. In classical stochastic processes, the observed time dependence of transition or decay proportional to $t^2$ is attributed to the noise with memory. In contrast to quantum mechanics, the quadratic form of initial decay is unable to decelerate the evolution of the system. 
\end{abstract}

  

\section{Introduction}
\label{Introduction} 

In many stochastic processes the survival probability $S(t)$ attracts considerable attention~\cite{Breuer,Weiss,Rivas,Fonda:1978dk,Bogdanowicz:1994bi}. The standard exponential decay law $S(t)=e^{-t/\tau}$  is never exact. For short times, $t\ll\tau$, the decay law is typically quadratic 
\begin{equation}
S\left(t\right) \simeq 1-\left(\frac{t}{\tau}\right)^2.
\end{equation}
Obviously a central part of such analysis is the use of the $t^2$ factor (or, at last in principle, $t^{\alpha}$ with $\alpha >1$) to produce a slowing down of decay by measurement. As a consequence, for frequently repeated measurements at very small time intervals, the Zeno effect, that is the freezing of state in its unstable initial configuration, takes place \cite{Misra:1976by,Koshino:2004rw,Facchi}. If $S\left(t\right)$ oscillates around exponential decay law, also the inverse Zeno effect, that is an increased decay rate by measurements, is possible~\cite{Facchi2}. Finally, at large times $t\gg\alpha\tau$ a power law sets in, $S(t) \simeq t^{-\alpha}$, where $\alpha > 0$ depends on the details of the interaction that rules that particular decay~\cite{Rothe}~\footnote{Non-exponential decay of unstable systems can also be understood and described in a most economical and adequate way in terms of non-extensive statistics \cite{Wilk:2001nv}}. It is the fact that $t^{2}-$ dependence for the quantum case leads to the quantum Zeno prediction, and should be regarded as the quantum mechanical norm. There is, thus, a fundamental mismatch between quantum physics and classical physics, both in the very specific point of the $t^{2}-$ dependence versus the $t-$ dependence for short time, and in the more general point that decay processes are exponential in classical case, but not in the quantum one. In this work we discuss classical processes influenced by noise with memory, which show $t^2$-form of initial decay, and we show that the Zeno effect vanishes in this case. 

In Section~\ref{sec:Quantum_Zeno_paradox} we recall the standard analysis of quantum decay. In section~\ref{sec:Non-local_master_equation_with_memory} we construct master equation with stochastic noise which solution for small enough $t$ show $t^{2}-$ form. Then, we discuss obtained solution (section~\ref{sec:Some_remarks}), and finally we advocate in section~\ref{sec:Memory destroy independence} that memory destroy independence and in a classical system an unstable particle observed continuously decay quasi-exponentially (the Zeno paradox does not exist). Section~\ref{sec:Summary} summarizes and concludes our work.


\section{Quantum Zeno paradox}
\label{sec:Quantum_Zeno_paradox}

Simple derivation of the quantum Zeno effect is possible by considering time behaviour of the state vector~\cite{Itano}. Let the state vector at time $t$ is $e^{-iHt}|\Phi\rangle$, where $|\Phi\rangle$ is the state vector at time $t=0$ and $H$ is the Hamiltonian in units where the reduced Planck's constant is set to unity, $ \hbar =1$.
For $t$ small enough, it is possible to make a power series expansion: $e^{-iHt}\simeq 1-iHt-H^2t^2/2+...$. The survival probability is
\begin{equation}
S(t)= | \langle\Phi|e^{-iHt}|\Phi\rangle|^2 \simeq 1-(\Delta H)^2t^2,
\label{eq:survival_q}
\end{equation}
where
\begin{equation}
(\Delta H)^2=\langle \Phi| H^2| \Phi\rangle-\langle\Phi| H| \Phi\rangle^2  
\label{eq:variance_H}
\end{equation}
is the variance of Hamiltonian.

Interrupted the interval $[0,t]$ by $n$ independent sequential measurements at times $t/n$, the survival probability is
\begin{equation}
S(t) \simeq \left [1-(\Delta H)^2 \left (\frac {t}{n}\right )^2 \right ]^n
\label{eq:mean}
\end{equation}
and approaches 1 for $n \rightarrow \infty$. 
It illustrates the situation that an unstable particle, if observed continuously, will never decay.
Zeno effect is a result of (a) a $t^2$-form of initial decay, which is the normal quantum case (cf. Eq.~(\ref{eq:survival_q})), and (b) the independence of sequential measurements (cf. Eq.~(\ref{eq:mean})) due to the fact, that in quantum mechanics, measurements generally change the state of the system being measured~\cite{Home}.


\section{Non-local master equation with memory}
\label{sec:Non-local_master_equation_with_memory}

In classical case we consider time-dependence of a state in circumstance where it may make transition to, or decay to, a second state. The rate of change of the non-decay probability $P(t)$ at time $t$ and dependence on its history (starting at $t=0$) is given by the non-local master equation
\begin{equation}
\frac{dP(t)}{dt}=-  \int_{0}^{t} \kappa(t-s)P(s) \, ds + \eta(t), 
\label{eq:master}
\end{equation}
where $\eta$ is the stochastic noise~\footnote{This approach to the evolution of statistical system is based on the Mori-Zwanzig formalism in which memory effects are taken into account through the introduction of the memory kernel $\kappa\left(t\right)$~\cite{Zwanzig}. This means that the rate of changes of the state $P\left(t\right)$ at time $t$ depends on its history (starting at $t=0$). Recently, the Langevin equation in approach given by Eq.~(\ref{eq:master}) was considered in description of particles motion~\cite{Ruggieri:2022kxv}}. The memory effects are taken into account through the introduction of the memory kernel 
\begin{equation}
\kappa(t,s)=\langle\eta(t) \eta(s)\rangle=\gamma f(t-s)
\label{eq:memory}
\end{equation}
with $\gamma = (\Delta \eta)^2 =\langle\eta ^2\rangle-\langle\eta\rangle^2 $ being variance of $\eta$ (notice that $\langle\eta\rangle=0$).
We assume that $\eta$ is a Lorentzian noise with spectrum
\begin{equation}
G(\omega)=\langle\eta^2\rangle\frac{\tau}{1+\omega^2 \tau^2}
\label{eq:Lorentzian}
\end{equation}
with correlators given by~\footnote{$\langle\eta(0)\eta(t)\rangle=\int_{0}^{\infty}{G(\omega)e^{i\omega t}d\omega}=\langle\eta^2\rangle e^{|t|/\tau}$}
\begin{equation}
f(t)=e^{-|t|/ \tau}
\label{eq:correlators}
\end{equation}
with $\tau$ being the memory time since it sets the time scale over which time correlations of noise decay~\footnote{For $f(t-s)=\delta(t-s)$ we have standard exponential decay law,  $S(t)=e^{-t/ \tau}$~\cite{Hanggi}}.

\begin{figure}[h]
\begin{center}
\resizebox{0.9\textwidth}{!}{
  \includegraphics{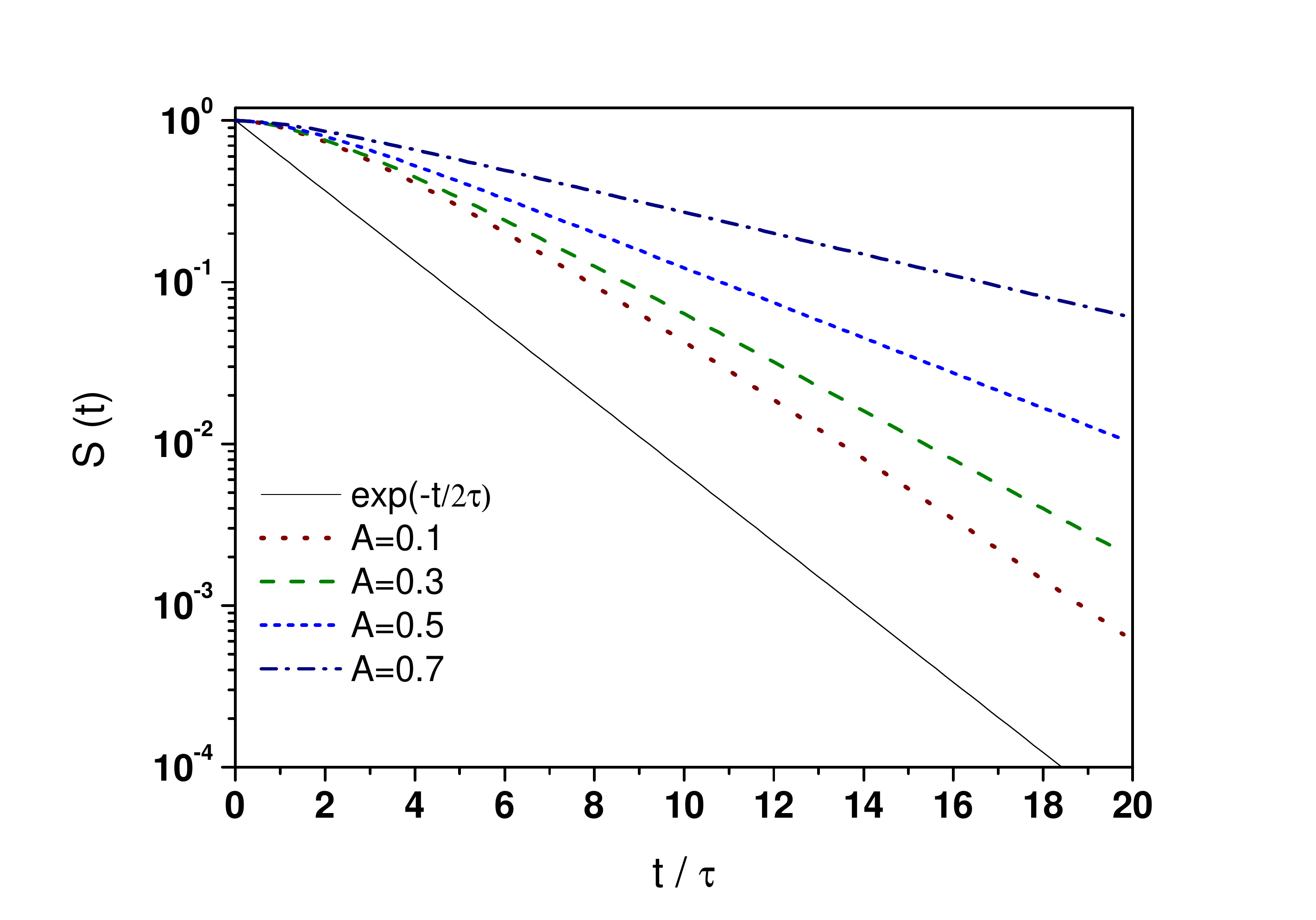}
  }
\caption{Survival probability $S(t)$ given by Eq.~(\ref{eq:final}) for different $A$ parameters.
 } 
\label{Fig_S}
\end{center}
\end{figure}

\begin{figure}[h]
\begin{center}
\resizebox{0.9\textwidth}{!}{
  \includegraphics{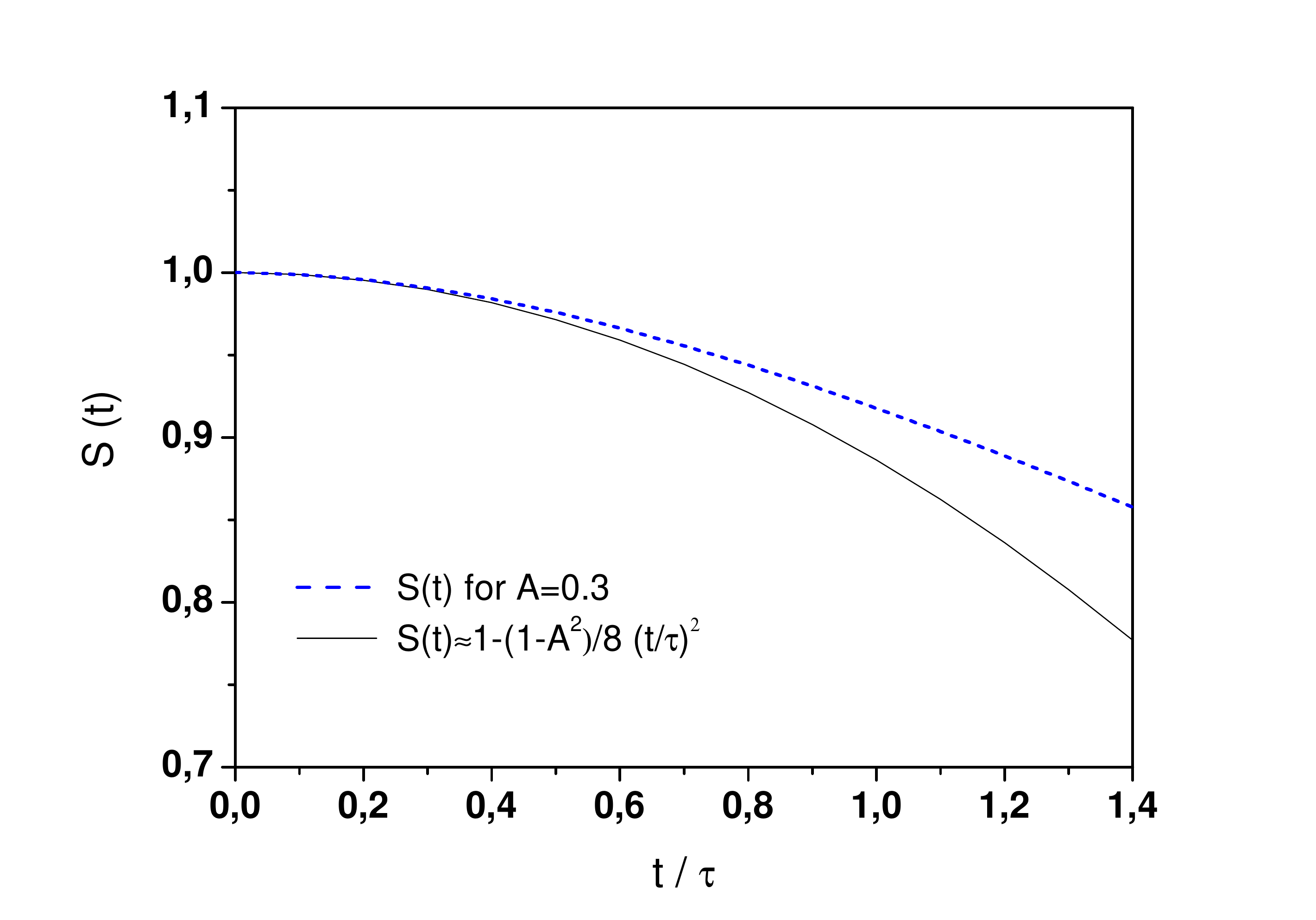}
  }
\caption{Initial time dependence of the survival probability $S(t)$ given by Eq.~(\ref{eq:final}) for parameter $A=0.3$ (broken line) in comparison with approximate parabolic behaviour given by Eq.~(\ref{eq:final_aprox}), with $\gamma=(1-A^2)/(4\tau^2)$ (solid line).
 } 
\label{Fig_S2}
\end{center}
\end{figure}

The non-decay probability $P\left(t\right)$ is by itself a random variable depending on the field fluctuations~\cite{Muller}. We define survival probability as the expectation value of the non-decay probability. Averaging probability over ensemble, for survival probability $S(t)=\langle P(t)\rangle$ we have
\begin{equation}
\frac{dS(t)}{dt}= - \int_{0}^{t} \gamma f(t-s) S(s) \, ds.
\label{eq:masterS}
\end{equation}
Differentiating both side of Eq.~(\ref{eq:masterS}) we have
\begin{equation}
\frac{d^2S(t)}{dt^2}+\frac{1}{\tau} \frac{dS(t)}{dt} +\gamma S(t)= 0
\label{eq:difS}
\end{equation}
which, with two initial conditions: 
$$
S\left(t=0\right)=0 \mbox {~and~} \frac{dS\left(t\right)}{dt}\Biggl\vert_{t=0}=0,
$$
allow simply to solve the problem analytically. Survival probability is given by
\begin{equation}
S(t) = \frac{e^{-t(1+A)/(2\tau)}(-1+A)}{2A}+ \frac{e^{-t(1-A)/(2\tau)}(1+A)}{2A}                 
\label{eq:final}
\end{equation}
with $A=\sqrt{1-4\gamma \tau ^2}$ (for formal solution of Eq.~\ref{eq:master} see Appendix \ref{sec:Appendix}).
Time dependence of the survival probability $S(t)$ for different parameters $A$ is shown in Fig.~(\ref {Fig_S}). Only for large $t$ we observe exponential behaviour,
$S(t) \sim \exp[-t (1-A)/(2 \tau)]$.
For $t$ small enough we have
\begin{equation}
S(t) \simeq 1-\frac{1}{2} \gamma t^2 = 1- \frac{1}{2 } ( \Delta \eta)^2 t^2       
\label{eq:final_aprox}
\end{equation}
with the same behaviour as $S(t)$ given by Eq.~(\ref{eq:survival_q}). Comparison of parabolic dependence given by Eq.~(\ref{eq:final_aprox}) with initial time dependence of $S\left(t\right)$ is shown in Fig.~(\ref{Fig_S2}). The variance of noise, characterized by $(\Delta \eta)^2 /2$ term, corresponds to $(\Delta H)^2$, and survival probability $S(t)$ the same exhibit $1-t^2$ form~\cite {Muller}.


\section{Some remarks}
\label{sec:Some_remarks}

Usually, the memory kernel $K(t-s)$  can lead to non-Markovian dynamics depending on the structure and time scale of the processes. Also a Markovian stochastic processes (where the time correlation $\langle\eta(t) \eta(s)\rangle \sim e^{-|t-s|/\tau}$ as is given by Eq.~(\ref {eq:memory})) can lead to non-Markovian dynamics of the system that is coupled to. In general, Markovianity or non-Markovianity are not just features of the noise but of the dynamics of the system coupled to the noise~\cite{Muller}.

Survival probability given by Eq.~(\ref{eq:final}) formally is the superposition 
\begin{equation}
S\left(t\right) = \omega_a \exp\left(-t/\tau_a\right)-\omega_b \exp\left(-t/\tau_b\right),
\label{eq:formal}
\end{equation}
with weights  $\omega_{a}=\left(1+A\right)/\left(2A\right)$, $\omega_{b}=\left(1-A\right)/\left(2A\right)$, and time constants $\tau_{a}=\left(2\tau\right)/\left(1-A\right)$, and $\tau_{b}=2\tau\left(1+A\right)$.
For $t$ small enough we have
\begin{equation}
S\left(t\right) \simeq \left(\omega_a-\omega_b\right)+\left(\frac{\omega_b}{\tau_b} - \frac{\omega_a}{\tau_a}\right) t +\left(\frac{\omega_a}{2 \tau^2_a} - \frac{\omega_b}{2\tau^2_b } \right) t^2.
\label{eq:formal_aprox}
\end{equation}
From normalization, $S(t=0)=1$, we have  $\omega_a-\omega_b=1$. For $\omega_a/\tau_a=\omega_b/\tau_b$ and $ \tau_a > \tau_b$ we observe $t^2$-form od initial decay (as given by Eq.~(\ref{eq:final_aprox})). However, for $\omega_a/\tau^2_a=\omega_b/\tau^2_b$ and $ \tau_a < \tau_b$ we have ``normal'' behaviour with $S(t) \simeq 1-(\omega_b/\tau_b-\omega_a/\tau_a) t$. 

The above discussion applies to $4\gamma \tau^{2}<1$ cases, nevertheless $4\gamma \tau^{2}>1$ is not excluded. Differential equation (\ref{eq:difS}) corresponds to damped harmonic oscillator. For $4\gamma \tau^{2}>1$ the system oscillate with amplitude gradually decreasing to zero (under-damped oscillator). The frequency of oscillations is given by $\omega^{2}=\gamma - \left(4\tau^{2}\right)^{-1}$ and amplitude decay with rate $1/2\tau$.~\footnote{Exponentially damped oscillating modes in decay of unstable quantum systems were widely discussed in~\cite{Giraldi:2019ymn,Winter:1961zz,RamirezJimenez:2021pok}.}


\section{Memory destroy independence}
\label{sec:Memory destroy independence}

Consider the survival probability $S(t) \simeq 1-at^2$, as given by Eqs.~(\ref{eq:survival_q}) and (\ref{eq:final_aprox}), where the time interval $[0,t]$ is interrupted by $n$  measurements at times $t/n, 2t/n,...,t$. The width of bins is $\delta=t/n$. Conditional probability for $i-th$ bin is
\begin{eqnarray}
 S_i=S\left(t_i=i\delta | t_{i-1}=(i-1) \delta \right) =\frac{S(i\delta \cap (i-1)\delta)}{S((i-1)\delta)}=
\nonumber 
\\  =\frac{S(i\delta)}{S((i-1)\delta)}=
\frac{1-a\delta^2i^2}{1-a\delta^2(i-1)^2}   \ \ \ \ \ \ \ \ \ \ \ \ \ \ \ \ \ \ \ \  \ \ 
\label{eq:S_i}
\end{eqnarray}
and is not the same as $S(\delta)=1-a\delta^2$, contrary to the memory-less exponential $S(t)$ for which $S_i=S(\delta)=e^{-\delta}$.
For $\delta \rightarrow 0$, the survival probability in each one of bins $S_i \rightarrow 1$.
Nevertheless, the survival probability for $n$ measurements
\begin{equation}
S(t=n\delta)=\prod_{i=1}^n \frac{1-a\delta^2i^2}{1-a\delta^2(i-1)^2}=1-at^2
\label{eq:S_n}
\end{equation}
and is not dependent on the number of measurements (in particular, $S(n\delta)$ is the same when $n \rightarrow \infty$). Contrary to arguments justifying the quantum Zeno effect, in classical system  an unstable particle observed continuously decay ``normally''.  

Exponential decay law, being memory-less,  obeys for conditional survival probability $S(t|t_1)$ the relation $S(t=t_1+t_2|t_1)=S(t_2)$. Due to independence we have $S(t_1+t_2)=S(t_1)S(t_2)$ (in particular, such assumption leads to Eq.~(\ref{eq:mean}) ).
This is not valid for dependent variables, and 
\begin{equation}
S(t=t_1+t_2) \neq S(t_1)S(t_2).
\label{eq:sum1}
\end{equation}
For dependent variables (due to noise correlation) we have chosen to use the joint survival probability, in the form (in analogy to the power expansion of multivariate normal distribution)
\begin{equation}
S(t_1,t_2,...,t_n)=  1-\sum^n_{i=1} \sum^n_{j=1} t_it_jC_{i,j},
\label{eq:joint_S}
\end{equation}
where $C_{i,j}=\langle\eta(t_i) \eta(t_j)\rangle-\langle\eta(t_i)\rangle\langle\eta(t_j)\rangle$ and for noise $\eta (t)$ we can write $C_{i,j}=C$.
Even for non-correlated noise, $C_{i,j}=0$ for $i \neq j$, we have $S(t_1,t_2,...,t_n)=1-\sum^n_i t^2_i C_{i,i} \neq S( \sum ^n_i t_t)$.

Interrupted the interval $[0,t]$ by $n$ measurements at times $\delta=t/n$, the survival probability is
\begin{equation}
S(\delta_1,\delta_2,...,\delta_n)=  1-C\sum^n_{i=1} \sum^n_{j=1} \left( \frac{t}{n} \right)^2  =1-Ct^2         
\label{eq:joint_delta}
\end{equation}
and the Zeno paradox does not exist~\footnote{The absence of Zeno effect in dynamics with classical evolution was considered recently in Ref.~\cite{Lopez:2020yjy}}.


\section{Summary}
\label{sec:Summary}

The dynamical evolution of statistical system is always influenced by their environment which exhibits time-correlated random fluctuations which can lead to non-Markovian dynamics~\cite{Benedetti1,Benedetti2}. By studying time-correlated noise we have shown how survival probability depends on the time scale of the noise correlations. The time correlations in the noise field determine how fast the survival probability converges to its exponential law behaviour. Approximations, considering only one variable survival probability $S(t)$ do not take into account all memory effects. The assumption of $S\left(\sum_{i=1}^n t_i \right)=\prod_{i=1}^nS(t_i)$, that  leads to the conclusion expressed by Eq.~(\ref{eq:mean}), is not valid in classical physics. Independence of sequential measurements in quantum mechanics arises due to the fact that, unlike in classical physics, measurements generally cannot be done without changing the state of the measured system. In contrast, in classical physics where measurements can be carried without affecting the state of the system, there is no Zeno effect even for quadratic form of the survival probabilities. 

\bigskip

This research was supported by the National Science Centre, Poland (NCN) Grant 2020/39/O/ST2/00277 (MR). In preparation of this work we used the resources of the Center for Computation and Computational Modeling of the Faculty of Exact and Natural Sciences of the Jan Kochanowski University of Kielce.
 

\appendix

\section{Formal solution of Eq.~(5)}
\label{sec:Appendix}

The formal solution of Eq.~(\ref{eq:master}) can be obtained by means of Laplace transform, namely:
\begin{equation}
P\left(t\right)=\frac{1}{2\pi\imath}\int_{\sigma-\imath\infty}^{\sigma+\imath\infty} \frac{P_{0}+N\left(s\right)}{s+K\left(s\right)}e^{st}ds,
\label{eq:A1}
\end{equation}
where $P_{0}=P\left(t=0\right)$, and $K\left(s\right)$ and $N\left(s\right)$ denote the Laplace transforms of the memory kernel and the noise, respectively. The integral is understood on a Bromwich contour, where $\sigma$ is a vertical contour in the complex plain chosen so that all singularities are to the left of it. 

We can evaluate the following averages:
\begin{equation}
S\left(t\right)=\langle P\left(t\right)\rangle=P_{0}\frac{1}{2\pi\imath}\int_{\sigma-\imath\infty}^{\sigma+\imath\infty} \frac{1}{s+K\left(s\right)}e^{st}ds,
\label{eq:A2}
\end{equation}
and
\begin{equation}
\langle \left(P\left(t\right)-\langle P\left(t\right)\rangle\right)^{2}\rangle=L^{-1}\left[\frac{\langle N\left(s\right)N\left(s'\right)\rangle}{\left(s+K\left(s\right)\right)\left(s'+K\left(s'\right)\right)}\right] \left(t,t'\right),
\label{eq:A3}
\end{equation}
where $L^{-1}\left[h\right]\left(t,t'\right)$ is the two-dimensional inverse Laplece transform of $h\left(s,s'\right)$ that depend on $t$ and $t'$. The Laplace transform of the memory kernel, given by Eq.~(\ref{eq:memory}) with noise correlator (\ref{eq:correlators}), is
\begin{equation}
K\left(t\right)=\frac{\tau\gamma}{\tau s+1},
\label{eq:A4}
\end{equation}
and
\begin{equation}
\langle N\left(s\right)N\left(s'\right)\rangle=\frac{\gamma\left(2+\tau s+\tau s'\right)}{\left(s+s'\right)\left(1+\tau s\right)\left(1+\tau s'\right)}.
\label{eq:A5}
\end{equation}
From Eqs.~(\ref{eq:A2}) and (\ref{eq:A4}), considering that the zeros of $s+K\left(s\right)$ are the solutions of the equation $s+\tau s^{2}+\gamma=0$, we get Eq.~(\ref{eq:final}) by a straightforward application of the residues theorem.

\end{document}